\documentclass[copyright,creativecommons,noderivs,noncommercial]{eptcs}

\usepackage{breakurl}
\usepackage{graphicx}

\begin{document}

\title{Implementing an Automatic Differentiator in ACL2}

\author{
  Peter Reid \\
      University of Oklahoma\\
	  Norman, Oklahoma, USA\\
      \email{peter.d.reid@gmail.com}
\and
  Ruben   Gamboa \\
      {Department of Computer Science}\\
      {University of Wyoming}\\
      {Laramie, Wyoming, USA}\\
      \email{ruben@cs.uwyo.edu}}

\def\authorrunning{P. Reid \& R. Gamboa}
\def\titlerunning{Implementing an Automatic Differentiator in ACL2}

\maketitle

\begin{abstract}
The foundational theory of differentiation was developed as
part of the original release of ACL2(r). In work reported at
the last ACL2 Workshop, we presented theorems justifying the
usual differentiation rules, including the chain rule and the
derivative of inverse functions. However, the process of
applying these theorems to formalize the derivative of a
particular function is completely manual.  More recently, we
developed a macro and supporting functions that can automate
this process.  This macro uses the ACL2 table facility to
keep track of functions and their derivatives, and it also
interacts with the macro that introduces inverse functions in
ACL2(r), so that their derivatives can also be automated. In
this paper, we present the implementation of this macro and
related functions.
\end{abstract}

\section{Introduction}

This paper describes the implementation of an automatic differentiator (AD)~\cite{ReGa:automatic-differentiator} that can 
find and prove the derivative of algebraic expressions in ACL2(r).  The tool is accessed through the macros
\texttt{defderivative} and \texttt{derivative-hyps}.  We will describe these macros more fully later,
but for now, we introduce them with an example.  

Suppose we have
defined the function \texttt{square} that computes $x \cdot x$.  
The following event determines the derivative of the function \texttt{square} and
proves the associated theorems:
\begin{verbatim}
    (defderivative square-deriv-local (square x))
\end{verbatim}
The key theorem that the macro \texttt{defderivative} introduces is as follows:
\begin{verbatim}
    (defthm square-deriv-local
      (implies (and (acl2-numberp x)
                    (acl2-numberp y)
                    (standardp x)
                    (i-close x y)
                    (not (equal x y)))
               (i-close (/ (- (square x) 
                              (square y)) 
                           (- x y))
                        (+ (* x 1) (* x 1))))
      :hints ...)
\end{verbatim}
The conclusion of this theorem states that the derivative of square is
        $x\cdot 1 + x\cdot 1$, which of course simplifies to the
        expected value of $2x$.  The macro does not automatically
        perform such simplifications, but the user can easily provide
        the preferred form for the derivative function\footnote{As we
        write this, we are working on a version of the macro that lets
        the user provide this function when the macro is introduced.
        This will simplify the process described here.}.  For example, 
        the function \texttt{square-prime} can be defined as \texttt{(* 2 x)}, 
        and the derivative theorems for it can be proved using the
        macro \texttt{derivative-hyps}:
\begin{verbatim}
    (derivative-hyps square
     :close-hints 
       (("Goal" 
         :use ((:instance square-deriv-local)) 
         :in-theory (disable square-deriv-local))))
\end{verbatim}  
The macro  \texttt{derivative-hyps} introduces all the theorems that
establish that the function \texttt{square-prime} is in fact the
derivative of \texttt{square}.  The keyword arguments allow the user
to provide hints for some of the necessary theorems.  In this case,
it is necessary to explicitly invoke the theorem
\texttt{square-deriv-local}, which was previously introduced via
\texttt{defderivative}.

The rest of this paper describes the implementation of
these macros.  Section~\ref{differences} describes some differences
in the theory of differentiation that proved useful in developing the
AD macros.  In particular, the proofs of the algebraic composition
theorems differ from the ones described in~\cite{GaCo:chain-rule}
to take advantage of the fact the derivative is known in the current
context.  Section~\ref{capabilities} describes the capabilities and
limitations of our AD system.  This is followed in 
Section~\ref{implementation} with a full description of their
implementation.  Finally, Section~\ref{conclusions} describes future
enhancements to these macros.

\section{The Revised Story of Differentiation in ACL2(r)}
\label{differences}

The theory of differentiation that was developed 
in~\cite{Gam:continuity} and~\cite{GaCo:chain-rule} is strictly
foundational.  For one thing, the development is concerned more
with differentiability than with derivatives. For example, the 
theorem that describes the derivative of sums is stated informally
as follows: If $f$ and $g$ are differentiable functions, so is $f+g$.
Notice that no mention is made of the derivatives of $f$, $g$, or
$f+g$!  Instead, the theorems deal directly with expressions
corresponding to the differential of the functions, e.g., 
$\Delta f(x)/\Delta x = (f(x+\epsilon)-f(x))/\epsilon$.

Using principles from non-standard 
analysis~\cite{Robinson:nsa,Robert:nsa}, these derivatives 
can be introduced implicitly by taking standard parts.  That is,
$f'(x)={}^*(\Delta f(x)/\Delta x)$.  However, this definition
only works when $x$ is standard.  It can be generalized using
\texttt{defthm-std}, but this process is unsatisfactory because
the relationship between $f'$, the expected derivative of the function $f$, and 
the standard part of $\Delta f/\Delta x$ is obscured.  This may explain why previous
results included many abstract theorems about differentiable
functions, but few concrete derivatives.  For instance, the
derivatives of the trigonometric, exponential, and logarithmic
functions had not been proven in ACL2(r) before this project.

A more significant challenge is the use of intervals in the
formalization of differentiation.  Intervals were used to define
the domain (and sometimes range) of differentiable functions.
This corresponds to typical mathematical statements.  For example,
one of the hypotheses of the mean-value theorem (MVT) is that $f$ 
is continuous over an interval $[a,b]$ and differentiable over
$(a,b)$, and a user can select suitable values of $a$ and $b$
when the theorem is applied manually.  But this is harder to do
when the theorems are applied automatically, if for no other
reason than the domain of some functions (such as $f(x)=1/x$)
is not a simple interval.

Consequently, we developed new versions of the differentiability
criterion that make explicit use of the differentiable function 
$f$ and its derivative $f'$.  We also redeveloped versions of the
composition theorems, namely
\begin{itemize}
  \item $(f+g)'(x)       = f'(x) + g'(x)$.
  \item $(f\cdot g)'(x)  = f(x)g'(x) + f'(x)g(x)$.
  \item $(f \circ g)'(x) = f'(g(x)) g'(x)$.
  \item $(f^{-1})'(x)    = 1/f'(f^{-1}(x))$, where $f^{-1}$ is the (compositional)
        inverse of $f$.
\end{itemize}
These new versions represent the domain (and sometimes range) of $f$ 
explicit via functions instead of intervals or some other data
structure.  The association between the function $f$, its derivative
$f'$ and its domain is kept using a naming convention; i.e.,
\texttt{f}, \texttt{f-prime}, and \texttt{f-domain-p}.

Using these new formalizations and the associated naming conventions 
was the key to automating the application of the algebraic composition
rules first formalized in~\cite{GaCo:chain-rule}.  However, the new
formalization does have some drawbacks.  First, there is no guarantee
that the domains used are, or even contain non-trivial intervals.
This means, for example, that the derivative of $|x|$ could be
vacuously formalized on the domain $x\in\{0\}$.  More seriously,
however, it prevents the application of the more foundational theorems
established earlier, such as the MVT.  We are investigating ways to
bridge our current work with prior results to remedy this issue.

The final, significant challenge is that some of the results obtained 
previously were proven in contexts that turned out to be too restrictive.  
Specifically, important theorems, such as the chain rule, which is used 
repeatedly during automatic differentiation, was developed only for 
real-valued functions.  However, the trigonometric functions in ACL2(r), 
such as sine and cosine, are defined in terms of the complex exponential 
function.  For example, $\sin(x) \equiv (e^{ix}-e^{-ix}) / 2i$.  So we 
developed a new formalization of the chain rule, which works for complex 
numbers.

\section{The Automatic Differentiation Macros}
\label{capabilities}

Previously, we discussed how the macros \texttt{defderivative} and
\texttt{derivative-hyps} to introduce the derivative of a function.
In this section, we will explore these and other related macros more
fully.

The macro \texttt{derivative-hyps} is used to generate automatically
the theorems required to show that the function \texttt{f-prime}
is the derivative of \texttt{f}.
The theorems and the associated naming conventions are as follows:
\begin{itemize}
\item \textbf{f-number}
\begin{verbatim}
(implies (f-domain-p x)
         (acl2-numberp (f x)))
\end{verbatim}  
\item \textbf{f-standard}
\begin{verbatim}
(implies (and (standardp x)
              (f-domain-p x))
         (standardp (f x)))
\end{verbatim}          
\item \textbf{f-continuous}
\begin{verbatim}
(implies (and (f-domain-p x)
              (standardp x)
              (f-domain-p y)
              (i-close x y))
         (i-close (f x) (f y)))
\end{verbatim}          
\item \textbf{f-prime-number}
\begin{verbatim}
(implies (f-domain-p x)
         (acl2-numberp (f-prime x)))
\end{verbatim}          
\item \textbf{f-prime-standard}
\begin{verbatim}
(implies (and (standardp x)
              (f-domain-p x))
         (standardp (f-prime x)))
\end{verbatim}          
\item \textbf{f-prime-continuous}
\begin{verbatim}
(implies (and (f-domain-p x)
              (standardp x)
              (f-domain-p y)
              (i-close x y))
         (i-close (f-prime x) (f-prime y)))
\end{verbatim}          
\item \textbf{f-close}
\begin{verbatim}
(implies (and (f-domain-p x)
              (standardp x)
              (f-domain-p y)
              (i-close x y)
              (not (equal x y)))
         (i-close (/ (- (f x) (f y))
                     (- x y))
                  (f-prime x)))
\end{verbatim}          
\end{itemize}   
These theorems are precisely the ones that will be used to
establish the constraints of \texttt{encapsulate}s that
are used to encode the composition theorems.  The names of
the theorems are important, because the macros will generate
hints with those names.

Similarly, the macro \texttt{inverse-hyps} generates the
theorems that establish that the function \texttt{f-inverse}
is the compositional inverse of \texttt{f}.
\begin{itemize}
\item \textbf{f-inverse-in-range}
\begin{verbatim}
(implies (f-inverse-domain-p x)
         (f-domain-p (f-inverse x)))
\end{verbatim}  
\item \textbf{f-domain-is-number}
\begin{verbatim}
(implies (f-domain-p x)
         (acl2-numberp x))
\end{verbatim}  
\item \textbf{f-inverse-relation}
\begin{verbatim}
(implies (f-inverse-domain-p x)
         (equal (f (f-inverse x)) x))
\end{verbatim}  
\item \textbf{f-d/dx-f-relation}
\begin{verbatim}
(implies (f-inverse-domain-p x)
         (equal (f-inverse-prime x)
                (/ (f-prime (f-inverse x)))))
\end{verbatim}  
\item \textbf{f-prime-not-zero}
\begin{verbatim}
(implies (f-domain-p x)
         (not (equal (f-prime x) 0)))
\end{verbatim}  
\item \textbf{f-preserves-not-close}
\begin{verbatim}
(implies (and (f-domain-p x)
              (f-domain-p y)
              (i-limited x)
              (not (i-close x y)))
         (not (i-close (f x) (f y))))
\end{verbatim}  
\end{itemize}
The theorems generated by this macro are precisely the constraints 
needed to establish that
the differentiable function $f$ has an inverse.  Note, for example,
the last two theorems above.  The theorem \textbf{f-prime-not-zero}
ensures that $f'(x)\ne0$ in the domain of $f$.  This is, in fact,
one of the hypotheses of the theorem that states that $(f^{-1})'(x) = 
1/f'(f^{-1}(x))$, since the expression contains $1/f'(\dots)$.

Both \texttt{defderivative} and \texttt{derivative-hyps} can be used
in two different contexts.  First, our own functions and macros use
them the generate constraints in various \texttt{encapsulate}s.
Second, the user can use these macros to generate the theorems that
correspond to these constraints, thus making sure that (a) the
constraints will be satisfied, and (b) the theorems satisfy the
naming conventions assumed by the macros.

One final point is related to the theorems generated by these macros:
These have to be proved by ACL2(r).  Many times, the proofs succeed
automatically, because the macros are careful to use the minimal
theory required for the proofs to go through, but sometimes ACL2(r)
needs a little help.  The macros use keyword arguments to accept hints 
that will be passed on to the appropriate generated theorems.  For 
example, the keyword argument \texttt{not-close-hints} is used to 
provide a hint to the theorem \texttt{f-preserves-not-close}.

The macro \texttt{defderivative} is the main entry point into the
automatic differentiator.  It takes two arguments, a prefix used to
scope the generated theorem names, and the arithmetic expression that
is to be derived.  By ``arithmetic expression'', we mean an ACL2 term 
that is composed only of numbers, variables, arithmetic operators, and
the application of functions with known derivatives or functions that 
are defined in terms of arithmetic expressions or as the inverse of 
functions that have known derivatives or are defined in terms of
arithmetic expressions.  In particular, recursive functions are not 
allowed, nor are functions that use \texttt{if} or \texttt{cond}.

What \texttt{defderivative} does is first to compute symbolically the
derivative of the expression, and then to generate the lemmas
necessary to demonstrate that the derivative computed symbolically
is correct.  Note that the lemmas follow the pattern and naming
convention defined by \texttt{derivative-hyps}, so the derivatives
of deeply nested expressions can be done automatically.

Finally, the macros \texttt{def-elem-derivative} and
\texttt{def-elem-inverse} are used to register functions with
known derivatives or inverses, respectively.  For example, the
AD system automatically defines the derivative of $f(x)=1/x$ with
the following event:
\begin{verbatim}
(def-elem-derivative 
  unary-/ 
  elem-unary-/ 
  (and (acl2-numberp x)
       (not (equal x 0)))
  (- (/ (* x x))))
\end{verbatim}  
The arguments to this macro are (1) the name of the function that 
has a known derivative (in this case, \texttt{unary-/}), (2) a prefix
used to name all the theorems generated by \texttt{derivative-hyps},
(3) an ACL2 expression for the domain of the function, and (4) an
ACL2 expression for the derivative.  Note that
\texttt{def-elem-derivative} does not prove any theorems.  Rather,
it registers in a global database that the given function has the
specified derivative.  It is expected that the proofs have been 
previously generated, e.g., using \texttt{derivative-hyps}.  

Similarly, \texttt{def-elem-inverse} is used to register an inverse
function.  For example, the following expression registers the
inverse of the function \texttt{square}:
\begin{verbatim}
(def-elem-inverse 
  square-inverse
  square-inverse
  (square-domain-p x)
  (square-inverse-domain-p x)
  square)
\end{verbatim}  
The arguments are similar to \texttt{def-elem-derivative}, except
both the domain and range (or inverse-domain) need to be specified.

\begin{figure}
\centering
\includegraphics[width=9.2cm]{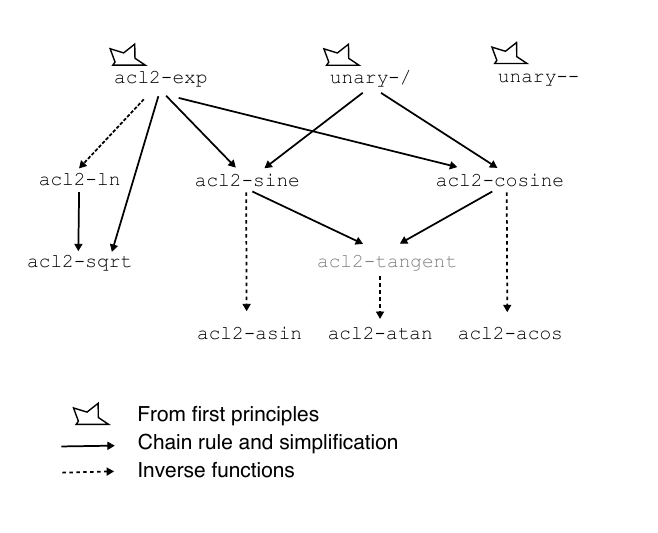}
\caption{Dependency graph of the functions built into \texttt{defderivative}. Symbols leading into a function represent how its derivative theorems were proved.}
\label{fig:dep-graph}
\end{figure}

As we mentioned previously, the AD automatically uses 
\texttt{def-elem-derivative} to register the derivatives of 
$f(x)=1/x$ and $f(x)=-x$.  This explains how subtraction and 
division are handled, namely through the chain rule and those
two derivative facts.  In addition, we have developed several
ACL2 books that establish the derivative of more complex functions,
such as $e^x$, $\ln x$, $\sin(x)$, $\sin^{-1}(x)$, etc.  The
derivative of $e^x$ was done using first principles, but the others
were done using the macros described in this section.  The
relevant books also register the derivatives of these functions 
using \texttt{def-elem-derivative}.  The derivative facts we have
established so far are summarized in Figure~\ref{fig:dep-graph}.
We note in passing that the macros support not just unary functions
but binary functions \emph{where one argument is held fixed}.  This
allows us to differentiate the \texttt{raise} function to find the
derivatives of $x^n$ and $a^x$.  The trick of holding an argument
fixed is essentially the same one used in~\cite{SaGa:sqrt}.

\section{Implementing the Macros}
\label{implementation}

The macros \texttt{derivative-hyps} and \texttt{inverse-hyps} simply
generate a \texttt{progn} containing several theorems.  There is not
much to their implementation.

Similarly, \texttt{def-elem-derivative} and \texttt{def-elem-inverse}
are nothing more than syntactic sugar for ACL2's built-in
\texttt{table} facility.  

That leaves the definition of \texttt{defderivative}.  In a nutshell,
this macro works by repeatedly applying the chain rule to
an expression, until each of its subexpressions is either a constant, a
variable, a function with a known derivative, or the inverse of a
function.  Before
describing this macro, we want to make a minor point.  Obviously,
\texttt{defderivative} needs access to ACL2's definition database,
so that it can expand function applications to compute derivatives.
However, access to these definitions depends on access to ACL2 
\texttt{state}, and ACL2 macros have traditionally not allowed 
access to \texttt{state}.  However, such access is now permitted 
via \texttt{make-event}\footnote{What \emph{did}
we do before \texttt{make-event}?}~\cite{make-event}.

The first thing that \texttt{defderivative} does is translate
the term to the derived, so that it does not contain any macros.
Among other things, this replaces terms using \texttt{+} with
terms using \texttt{binary-+}.  Then, the resulting expression
is differentiated symbolically.  During this process, the
necessary proofs are collected and laid out using
\texttt{encapsulate}.  Many of the proofs are done automatically
by instantiating the relevant composition theorems.  The macros
know the name of the composition theorems and their constrained
functions, so they can generate the appropriate hints.  Note
that this process also involves the symbolic computation of the 
domain of intermediate functions.

The proofs of these theorems need to be fully automated, since there
is no way for the user to give hints, as the theorems may be
associated with arbitrarily deep subterms of the original formula.  
We try to guarantee this automation by using only
minimal theories, usually only the names of the theorems generated
by \texttt{derivative-hyps} and \texttt{inverse-hyps}.  This is
one reason why users must conform to these naming conventions, even
when they prove a derivative fact from first principles, as we did
for $e^x$.

The last thing that \texttt{defderivative} does is to clean up the
expressions generated for the derivative and domain of the function.
This is, perhaps, the biggest weakness of our current implementation.
The clean-up process is simplistic, consisting mainly of converting
\texttt{binary-+} back to \texttt{+}.  We have considered invoking
ACL2's rewriter at this point, if only to perform arithmetic
simplification, e.g., to convert \texttt{(+ (* x 1) (* x 1))} to
\texttt{(+ x x)}.  But we have not found a satisfactory set of rewrite
rules, so we are leaving the sophisticated rewrites to the user.

\section{Conclusions}
\label{conclusions}

This paper described the implementation of an automatic
differentiation (AD) system for ACL2(r).  The implementation
brought up several points of interest to the ACL2 community.

First, the idea of using macros to generate theorems according
to some pattern is as old as ACL2 (at least).  However, the
current work shows how different macros can cooperate by
keeping information in the ACL2 \texttt{state}.  For example,
the macro \texttt{definv} can register information about inverse
functions, which is subsequently used by the macros
\texttt{defderivative}.

Second, since ACL2 \texttt{state} is now available to macros,
the macros can generate code that depends on the ACL2
database.  For instance, the macros can use the definitions
of ACL2 functions to generate theorems according to some
pattern.

Third, our approach demonstrates the care that must be taken
when designing libraries intended for automation.  The history
of ACL2 and the Boyer-Moore theorem prover includes several
examples of libraries that are carefully designed so that new
theorems can be proved almost automatically.  The lemmas in
these libraries are chosen carefully so that they work well with
ACL2's heuristics (and vice versa).  But when a macro develops
a complex theory from arbitrary ACL2 expressions, it becomes
increasingly likely that some rewrite rules triggered by the ACL2
expression interfere with the proof plan of the macro.  So the
macro has to take careful control of the proof execution, especially
if hints are involved to instantiate constrained functions.  In
our experience, we have improved our chances by explicitly
controlling the active ACL2 theory, and making sure it only
has the rewrite rules that we think are absolutely necessary for
the theorems to prove.  We tried to use \texttt{:by} hints in
these instantiations, so that the proof plan was completely
controlled by the hints.  Unfortunately, we ran into too many
cases where \texttt{lambda} expressions created for derivatives
did not exactly match the functional instantiation generated
with a \texttt{:by} hint, so we had to switch to regular
\texttt{:use} hints instead.  So far, the proof plans are
succeeding, but we would prefer to have a more robust mechanism.

Fourth, our solution illustrates how naming conventions can 
be used to associate facts with functions.  For example, the
fact that \texttt{square} is 1-to-1 in its domain can be stored
in the theorem \texttt{square-is-1-to-1} and the domain
itself in the function \texttt{square-domain-p}.  Such conventions
enable macros to generate appropriate hints.  Moreover, 
maintaining those conventions is easy if the macros themselves
generate the required theorems.  Things are more complicated
when some of the theorems need to be generated by hand, e.g.,
to show the derivative of $e^x$.

Finally, the techniques described in this paper have enabled us
to vastly extend the derivative facts that have been certified
with ACL2(r).  As part of this project, we demonstrated from first
principles that $d e^x/dx = e^x$.  Then we used the macros that
we developed to formalize the derivatives of the trigonometric
functions, and the inverse trigonometric and logarithmic functions.

\bibliographystyle{alpha}
\bibliography{rag}

\end{document}